\documentclass[aps,pra,twocolumn,showpacs]{revtex4-1}
\usepackage{dcolumn}
\usepackage{multirow}
\begin{document}
[Phys. Rev. A {\bf 82}, 023603 (2010)]
\title{Ground States of a Mixture of Two Species of
Spinor Bose Gases with Interspecies Spin Exchange
}
\author{Yu Shi}
\email{yushi@fudan.edu.cn}
\affiliation{Department of Physics, Fudan University, Shanghai 200433, China}
\date{Received by Phys. Rev. A on 11 December 2009}
\begin{abstract}
We consider a mixture of two species of spin-1 atoms with
interspecies spin exchange, which  may  cooperate or compete with
the intraspecies spin exchanges and thus dramatically affect the
ground state. It represents a new class of bosonic gases differing
from  single-species spinor gases.  We determine the exact ground
states in several parameter regimes, and study the composite
structures by using  the generating function method generalized here
to be applicable to a mixture of two species of spinor gases. The
most interesting phase is the so-called entangled Bose-Einstein
condensation (BEC), which is fragmented BEC with quantum entanglement
between the two species, and with both interspecies and intraspecies
singlet pairs. For comparison, we also apply the generating function
method to a mixture of two species of pseudospin-$\frac{1}{2}$
atoms, for which the total spin quantum number  of each species is
fixed as half of the atom number, in contrast with the case of
spin-$1$, for which it is a variable determined by energetics.
Consequently, singlet pairs in entangled BEC of a
pseudospin-$\frac{1}{2}$ mixture are all interspecies. Interspecies
spin exchange leads to novel features beyond those of spinor BEC
of a single species of atoms as well as mixtures without
interspecies spin exchange.
\end{abstract}


\maketitle

\section{Introduction}

Spin-exchange scattering between bosonic atoms leads to novel
ground states and interesting phenomena unexplored
previously~\cite{ho1,law,koashi,ho2,spinor}. This remarkable subject
broadens the scope of magnetism, which is traditionally based on
spin exchanges of fermions instead. Mixtures of two
species without a spin degree of freedom or, equivalently, mixtures of
atoms  of the same species with two spin states whose occupation
numbers are both conserved have also been
studied~\cite{two}. What about a mixture of two different species
with interspecies spin exchange, in addition to  intraspecies
spin exchanges? This question was first explored in a mixture of two
species of pseudospin-$\frac{1}{2}$
atoms~\cite{shi0,shi1,shi2,shi4}, where the most interesting phase
was found to be BEC of interspecies singlet pairs, which was called
entangled Bose-Einstein condensation (EBEC), or BEC with an
entangled order parameter, emphasizing the aspect that it is in an
entangled state of two distinguishable atoms that BEC occurs. More
generally, here we define EBEC as multispecies BEC with interspecies
entanglement. It is a special kind of fragmented BEC. In EBEC,
quantum entanglement, which is the most essential quantum feature
not existing in classical physics, is amplified to a macroscopic
quantum phase, just as BEC in a superposed single-particle
state amplifies single-particle  superposition to a macroscopic
phase, leading to the Josephson effect.  EBEC also bears some
similarities to the lowest energy state of a SU(2) symmetric
model of a single species of  pseudospin-$\frac{1}{2}$ atoms which
are forced to occupy two orbital modes by conservation of the total
spin in the cooling process~\cite{kuklov}, but there are also
differences~\cite{shi4}.

To pick out two hyperfine states of an alkali atom representing
pseudospin-$\frac{1}{2}$  avoiding spin-exchange loss to other
hyperfine states may require some careful tuning~\cite{shi4}. In addition, with the degree of freedom being pseudospin, there is no
reason to expect rotation symmetry in the interaction. In contrast,  in a mixture of two species of spin-$1$ atoms with full access
to the spin-$1$ multiplet, the total single-particle energy of any
two scattered particles is always conserved, whether in the absence or
the presence of a magnetic field, hence  spin-exchange scattering
is energetically protected. Moreover, the total spin of each species is
conserved by the intraspecies interaction, while the total spin of
the whole mixture is conserved by the interspecies interaction.
Therefore, if it is also a ground state of  a mixture of two species
of spin-$1$ atoms,  EBEC may be more experimentally accessible and
stable in such a mixture than in a pseudospin-$\frac{1}{2}$ mixture.
However, our consideration of spin-$1$ mixture had been impeded by
the apparent complexity of the spin-exchange interaction between two
spin-1 atoms of different species, until it was shown recently that
it is simply of Heisenberg form~\cite{ma}.  Some mean-field-like
investigations of spin-$1$ mixtures have been carried out, but possible
entanglement between the two species was ignored~\cite{ma,xu}.

In this article, we rigorously study the ground states of a mixture
of two species of spin-$1$ atoms in various parameter regimes.
Bosonic symmetry within each species and its absence between
different species together lead to rich structures, with interesting
features beyond those of a single species of spinor gas. For a
spin-$1$ mixture, we find EBEC in certain parameter regimes, with
the two species significantly entangled. Two atoms of the two
different species can form interspecies singlet pairs,  with
differences from a pseudospin-$\frac{1}{2}$ mixture, however; in EBEC
of a spin-$1$ mixture, intraspecies and interspecies singlet pairs
coexist.

The rest of the paper is organized as follows. The many-body Hamiltonian is given in Sec.~II. Under a common assumption for spin-1 bose gases, the present  Hamiltonian can be written solely in terms of spin operators of the two species and of the total system.  Then in Sec.~III, we find the ground states in terms of spin quantum numbers, in various parameter regimes. These ground states are given in terms of boson creation operators in Sec.~IV. In Sec.~V, composite structures of these ground states are studied by generalizing a generating function method from a single species to a mixture. In Sec.~VI, we discuss a mixture of two species of pseudospin-$\frac{1}{2}$  bosons. The paper is summarized in Sec.~VII.

\section{The Many-body Hamiltonian}

For two spin-$f$ atoms of different species, there is no permutation
symmetry between them, hence the total spin can be $F=0,\cdots 2f$.
The effective interaction is thus
\begin{eqnarray}
V(\mathbf{r}_a-\mathbf{r}_b) & = & \delta(\mathbf{r}_a-\mathbf{r}_b)\sum_{F=0}^{2f}g^{ab}_FP_F \\
&
= & \delta(\mathbf{r}_a-\mathbf{r}_b)\sum_j^{2f} \bar{c}^{ab}_{2j}
(\mathbf{F}_a \cdot \mathbf{F}_b)^j,
\end{eqnarray}
where $g^{ab}_F$ is the
interaction strength proportional to the $F$-channel scattering length, and
$P_F$ is the projection operator for the total spin $F$, and can be
expanded in terms of $1, \mathbf{F}_a\cdot
\mathbf{F}_b, \cdots, (\mathbf{F}_a\cdot
\mathbf{F}_b)^{2f}$. For $f=1$, $\bar{c}^{ab}_0 =
-g^{ab}_0/3+g^{ab}_1+g^{ab}_2/3$, $\bar{c}^{ab}_2 =
-g^{ab}_0/2+g^{ab}_2/2$, and $\bar{c}^{ab}_4 =
g^{ab}_0/3-g^{ab}_1/2+g^{ab}_2/6$. It has been shown that
$\bar{c}^{ab}_0=3 g_T/4+g_S/4$, $\bar{c}^{ab}_2=(g_T-g_S)/16$, and
$\bar{c}^{ab}_4=0$, where $g_T$ and $g_S$ correspond to the triplet
and singlet states of the two valence electrons of the scattering
atoms~\cite{ma}.

Therefore  the many-body Hamiltonian is
\begin{equation}
{\cal H} = {\cal H}_a + {\cal H}_b + {\cal H}_{ab},
\end{equation}
where
\begin{equation}
\begin{array}{l}
\displaystyle {\cal H}_{\alpha} = \int d\mathbf{r}
\psi^{\dagger}_{\alpha\mu}h_{\alpha} (\mathbf{r})_{\mu\nu}
\psi_{\alpha\nu} \\ \displaystyle + \frac{1}{2} \int d\mathbf{r}
\psi^{\dagger}_{\alpha\mu}\psi^{\dagger}_{\alpha\rho}
(\bar{c}_0^{\alpha}\delta_{\mu\nu}
\delta_{\rho\sigma}+\bar{c}_2^{\alpha}
\mathbf{F}_{\alpha\mu\nu}\cdot\mathbf{F}_{\alpha\rho\sigma})
\psi_{\alpha\sigma}\psi_{\alpha\nu}
\end{array}
\end{equation}
is the well-known Hamiltonian of spin-1 atoms~\cite{ho1} of species $\alpha$ ($\alpha = a, b$), while
\begin{equation}
{\cal H}_{ab}= \int d\mathbf{r}
\psi^{\dagger}_{a\mu}\psi^{\dagger}_{b\rho}
(\bar{c}_0^{ab}\delta_{\mu\nu} \delta_{\rho\sigma}+\bar{c}_2^{ab}
\mathbf{F}_{a\mu\nu}\cdot\mathbf{F}_{b\rho\sigma})
\psi_{b\sigma}\psi_{a\nu}
\end{equation}
is the interspecies interaction. Here the field operator $\psi_{\alpha\mu}$ corresponds
to spin $\mu$ component of species $\alpha$ ($\mu = -1, 0, 1$), and
$\mathbf{F}_{\alpha\mu\nu}$ represents the $(\mu\nu)$ element of the
spin-$1$ matrix of species $\alpha$.
\begin{equation}
h_{\alpha}=-\frac{\hbar^2}{2m_{\alpha}}\nabla^2+U_{\alpha} (\mathbf{r})
-\gamma_{\alpha}\mathbf{B} \cdot
\mathbf{F}_{\alpha}
\end{equation}
is the single particle Hamiltonian of species
$\alpha$, $m_{\alpha}$ and $\gamma_{\alpha}$ are the  mass and the
gyromagnetic ratio of an atom of species $\alpha$, respectively,  $\mathbf{B}$ is a
uniform magnetic field, and $U_{\alpha}$ is the trapping potential for an atom of species $\alpha$.

For the single-particle orbital wave function, certainly one can use the usual single-mode approximation to write $\psi_{\alpha\mu}(\mathbf{r})=
\alpha_{\mu}\phi_{\alpha\mu}(\mathbf{r})$, where $\alpha_{\mu}=a_{\mu},
b_{\mu}$ is the annihilation operator and  $\phi_{\alpha\mu}$ is the
lowest single-particle orbital wave function for species $\alpha$ and spin $\mu$. Then we obtain a Hamiltonian in terms of creation and annihilation  operators, with integrals of various products of  $\phi_{\alpha\mu}$'s  and their complex conjugates entering as coefficients of the terms of the Hamiltonian.

Nevertheless, to simplify the matter, we can follow an additional common assumption for spin-$1$ bose gases, namely that the single particle orbital wave function is
mainly determined by the spin-independent part of the Hamiltonian
and is thus independent of spin; that is, $\phi_{\alpha\mu}=\phi_{\alpha}$ is independent of spin $\mu$.
For a homogeneous system, $\phi_{\alpha}=1/\sqrt{\Omega}$, where
$\Omega$ is the volume. But our discussions also apply to the  inhomogeneous case.

Therefore, the Hamiltonian can be simplified as
\begin{equation}
{\cal H} = \frac{c^{a}}{2} \mathbf{S}_a^2 +
\frac{c^{b}}{2}\mathbf{S}_b^2 + c^{ab} \mathbf{S}_a \cdot
\mathbf{S}_b - \gamma_a \mathbf{B}\cdot \mathbf{S}_{a} - \gamma_b
\mathbf{B}\cdot \mathbf{S}_{b}, \label{spinham}
\end{equation}
where
\begin{equation}
\mathbf{S}_{\alpha}=
\alpha^{\dagger}_{\mu}\mathbf{F}_{\mu\nu}\alpha_{\nu}
\end{equation}
is the total spin operator for species  $\alpha$, $\gamma_a, \gamma_b > 0$,
$C=N_a\epsilon_a+N_b\epsilon_b +\frac{c_0^{a}}{2}(N_a^2-N_a)+
\frac{c_0^{b}}{2}(N_b^2-N_b) - c^{a}N_a - c^{b}N_b +
c_0^{ab}N_aN_b$ is a constant, $c_k^{\alpha}=\bar{c}_k^{\alpha}\int
d^3r |\phi_{\alpha}|^4$, $c_k^{ab}=\bar{c}_k^{ab}\int d^3 r
|\phi_{a}\phi_b|^2$, ($k=0,2$), and we have simplified notations  $c_2^{\alpha}$ and $c_2^{ab}$  as $c^{\alpha}$ and $c^{ab}$, respectively.

If $c^{ab}=0$, then there is no
entanglement between the two species of atoms, the ground state is
simply a direct product of the ground states of the two species of
spin-1 atoms.

We now set out to find the ground states of (\ref{spinham}) in various parameter regimes.

\section{Ground states in terms of spins}

We assume $\gamma_a=\gamma_b=\gamma \geq 0$, as satisfied by alkali
atoms with the same nuclear spin, for example,  $^7$Li, $^{23}$Na, $^{39}$K,
$^{41}$K and $^{87}$Rb,  all of which have nuclear spin $3/2$.

Then $S_a$ and $S_b$ together with the total spin $S$ and its
$z$-component $S_z$ are all conserved. Therefore, the ground state is
\begin{equation}
|G\rangle = |S_a,S_b,S,S_z\rangle,
\end{equation}
with the four spin quantum
numbers being the integers that minimize the energy
%\begin{equation}
$E=\frac{c^{a}-c^{ab}}{2 } S_a(S_a+1)+
\frac{c^{b}-c^{ab}}{2 }S_b(S_b+1)+ \frac{c^{ab}}{2 } S(S+1) -
\gamma B S_z,$
%\end{equation}
where $B>0$ is the magnitude of the magnetic field and
the constant $C$ has been neglected.
Note that the minimization of $E$ is
under the constraints $|S_a-S_b| \leq S \leq S_a+S_b$ and $-S\leq
S_z \leq S$.

Moreover, for a given $S$, $S_z=S$ minimizes the energy. Therefore the ground state must be
\begin{equation}
|G\rangle = |{\cal S}_a^m,{\cal S}_b^m,{\cal S}^m,
{\cal S}^m\rangle, \label{gs} \end{equation} where ${\cal S}_a^m$, ${\cal
S}_b^m$ and ${\cal S}^m$ are, respectively, the values of $S_a$,
$S_b$ and $S$ that minimize
\begin{widetext}
\begin{equation}
E=\frac{c^{a}-c^{ab}}{2 }
S_a(S_a+1)+ \frac{c^{b}-c^{ab}}{2 }S_b(S_b+1)+ \frac{c^{ab}}{2
} S(S+1) - \gamma B S. \label{e}
\end{equation}
\end{widetext}

In the following, we find the  ground states in the form of
(\ref{gs}), by minimizing (\ref{e}), for various parameter regimes.
For specification, we assume $N_a \geq N_b$ without loss of
generality.

\subsection{$c^{ab} < 0$}

First, we consider the cases with $c^{ab} < 0$. Then for given $S_a$
and $S_b$, it is $S=S_z=S_a+S_b$ that minimizes the energy (\ref{e}).

The ground state is thus
\begin{equation} |G(c^{ab} < 0)\rangle = |{\cal
S}_a^m,{\cal S}_b^m,{\cal S}_a^m+{\cal S}_b^m, {\cal S}_a^m+{\cal
S}_b^m\rangle,\end{equation} where ${\cal S}_a^m$ and ${\cal S}_b^m$
are respectively  the values of $S_a$ and $S_b$ that minimize
$E(c^{ab}<0)=\frac{c^{a}-c^{ab}}{2 } S_a(S_a+1)+ \frac{c^{b}-c^{ab}}{2
}S_b(S_b+1)+ \frac{c^{ab}}{2 } (S_a+S_b)(S_a+S_b+1) - \gamma B
(S_a+S_b)$, and must be calculated separately for different subcases.

In particular, here we consider $c^{ab}<0$,
$c^{a} < c^{ab}$ and $c^{b} < c^{ab}$.  $E$ always
decreases as $S_a$ or $S_b$ increases. For each species of spin-1
atoms, the largest possible value of the spin quantum number is
clearly its number of particle. Hence ${\cal S}_a^m=N_a$ and ${\cal
S}_b^m=N_b$. Therefore the ground state is
\begin{eqnarray}
|G\rangle_I & = & |N_a,N_b,N_a+N_b,N_a+N_b\rangle \\
& =
&¡¡|N_{a}, N_a\rangle_a\otimes |N_b, N_b\rangle_b,
\end{eqnarray}
where the subscript ``$I$'' in $|G\rangle_I$ denotes the parameter
regime. Similar notations are used for ground states in other
parameter regimes. Throughout this paper, a spin basis state written
without a subscript is of the total system, while a spin basis state
with a subscript ``$\alpha$'' is of species $\alpha$; that is,
$$|S_{\alpha},S_{\alpha z}\rangle_{\alpha}, $$
is a state of species $\alpha$ with total spin $S_{\alpha}$ and its
$z$ component $S_z$. Obviously $|G\rangle_I$ is the direct product
of the ferromagnetic states of the two species.

In the case of $c^{ab}<0$, $c^{a}<c^{ab}$ and
$c^{b}<c^{ab}$, if the magnetic field is absent, that is,  $B=0$,
then there are $2(N_a+N_b)+1$ degenerate ground states
\begin{widetext}
\begin{eqnarray}
|G\rangle_{I'}&=&|N_a,N_b,N_a+N_b,S_z\rangle\\ & =& \sum_{S_{bz}=-N_b}^{N_b}
g_{I'}(S_{bz}) |N_a, S_z-S_{bz}\rangle_a \otimes |N_b,
S_{bz}\rangle_b, \label{g12}
\end{eqnarray}
\end{widetext}
where $S_z=-N_a-N_b,\cdots, N_a+N_b$, the prime in the subscript to
$|G\rangle_{I'}$  represents the absence of a magnetic field, and $g_{I'}(S_{bz})$
is the Clebsch-Gordan coefficient.

With $c^{ab}<0$, one can see that the ground state remains as
$|G\rangle_I$ or $|G\rangle_{I'}$ if $c^{a}=c^{ab}$ while
$c^{b} < c^{ab}$, or $c^{b}=c^{ab}$ while $c^{a}
< c^{ab}$, or $c^{a}= c^{b} = c^{ab}$. This is
because to minimize $E$, first $S$ is maximized to the largest possible value $N_a+N_b$.
Consequently $S_a$ and $S_b$ have to be  maximized to  $N_a$ and $N_b$,
respectively, even though one or both disappear in  $E$. Therefore
regime I can be expanded as $c^{ab}<0$, $c^{a}\leq
c^{ab}$ and $c^{b} \leq c^{ab}$.

\subsection{$c^{ab}>0$}

Now we turn to  cases with  $c^{ab}>0$, for which it is useful
to rewrite $E$ as $E(c^{ab}>0) =\frac{c^{a}-c^{ab}}{2 } S_a(S_a+1)+
\frac{c^{b}-c^{ab}}{2 }S_b(S_b+1)+ \frac{c^{ab}}{2 }
(S+\frac{1}{2}-\frac{\gamma B
 } {c^{ab}})^2 -\frac{c^{ab}}{2 }(\frac{1}{2}-\frac{\gamma B  }
{c^{ab}})^2$.
When $ c^{ab} \geq 2\gamma B $,
$c^{a}-c^{ab}>0$, and $c^{b}-c^{ab}>0$, the ground state
is
\begin{eqnarray} |G\rangle_{II} &
= & |0,0,0,0\rangle  \label{gii1} \\
& = & |0,0\rangle_a\otimes |0,0\rangle_b, \label{gii2}
\end{eqnarray}
which is the direct product of the two singlet states of the two
species. The result is valid even when $ c^{ab} = 2\gamma B =0 $, as $S_a$ and $S_b$ are still minimized, respectively, although such a trivial case is not our focus.

If $c^{ab} \geq 2\gamma B
>0$ or $c^{ab} > 2\gamma B =0$, one can see that the ground state remains as $|G\rangle_{II}$
if $c^{a}>c^{ab}$ while $c^{b} = c^{ab}$, or
$c^{b}>c^{ab}$ while $c^{a}
=c^{ab}$. This is because
to minimize $E$, $S$ and $S_a$ or
$S_b$ are  minimized to  $0$ and, consequently,  $S_b$ or
$S_a$ has to be $0$ too, although it does not appear
in  $E$.

Now let us consider the case with $c^{ab} \geq 2 \gamma B
>0$ or $c^{ab} > 2\gamma B =0$,
together with  conditions $c^{a}-c^{ab}<0$, $c^{b}-c^{ab}<0$
and $N_a=N_b=N$. The ground state is then the global singlet
state
\begin{widetext}
\begin{eqnarray}
|G\rangle_{III} & = & |N,N,0,0\rangle   \\
\displaystyle  & =  & \frac{1}{2N+1}\sum_{m=-N}^N (-1)^m
|N,m\rangle_a\otimes|N,-m\rangle_b,
 \label{g31}
\end{eqnarray}
\end{widetext}
which is a maximally entangled
state, as one can easily see by considering the reduced density matrix of either species.

In general, for two species of
spin-$f$ atoms of an arbitrary $f$, with $N_a=N_b=N$, a singlet state with maximal $S_a$ and $S_b$ is
\begin{widetext}
\begin{equation}
|fN,fN,0,0\rangle = \frac{1}{2fN+1}\sum_{m=-fN}^{fN} (-1)^m
|fN,m\rangle_a \otimes |fN,-m\rangle_b, \label{f}
\end{equation}
\end{widetext}
which is a maximally entangled state.

If $c^{ab} \geq 2\gamma B
>0$ or $c^{ab} > 2\gamma B =0$ under the constraint $N_a=N_b=N$, one can see that the ground state remains as $|G\rangle_{III}$
if $c^{a}<c^{ab}$ while $c^{b} = c^{ab}$, or
$c^{b}<c^{ab}$ while $c^{a}
=c^{ab}$. This is because
to minimize $E$, $S$ is  minimized to  $0$ while  $S_a$ or
$S_b$ is maximized to $N$ and, consequently, the other $S_b$ or
$S_a$ is maximized to $N$ also, although it does not appear
in $E$.

If $c^{ab} \geq 2\gamma B
>0$ or $c^{ab} > 2\gamma B =0$, while  $c^{a}=c^{b} = c^{ab}$, then the ground state is degenerate, and is $|S_b,S_b,0,0\rangle$, with $0\leq S_b \leq N_b $.

Finally,  if  $0< c^{ab} \leq 2\gamma B  $, $c^{a}-c^{ab}<0$, and
$c^{b}-c^{ab}<0$, while $|N_a-N_b| \leq  n \leq N_a+N_b$, where $ n \equiv Int(\frac{\gamma B
}{c^{ab} }- \frac{1}{2})$, then $E$ is minimized when $S_a=N_a$,
$S_b=N_b$,  and $S=n$. Here $Int(x) $ represents the integer closest to $x$ and in the legitimate range given above.¡±is the integer no larger than and closest to  $x$.  Hence the ground state is
\begin{widetext}
\begin{eqnarray}
|G\rangle_{IV} & = & |N_a,N_b,n,n\rangle\\
& = & \sum_{S_{bz}=-N_b}^{N_b}
g_{IV}(S_{bz}) |N_a, n-S_{bz}\rangle_a \otimes |N_b,
S_{bz}\rangle_b, \label{g4}
\end{eqnarray}
\end{widetext}
where $g_{IV}(S_{bz})$ is the  Clebsch-Gordon coefficient.

If  $0< c^{ab} \leq 2\gamma B $,  and $c^{a}-c^{ab}<0$ while
$c^{b}=c^{ab}$, then the ground state is  $|N_a,S_b,n,n\rangle$, under the constraint $|N_a-S_b| \leq n \leq N_a+S_b$.  If  $0< c^{ab} \leq 2\gamma B $, and $c^{b}-c^{ab}<0$ while
$c^{a}=c^{ab}$, then the ground state is  $|S_a,N_b,n,n\rangle$, under the constraint $|S_a-N_b| \leq n \leq S_a+N_b$. If $0< c^{ab} \leq 2\gamma B $ while  $c^{a}=c^{b}=c^{ab}$, then the ground state is  $|S_a,S_b,n,n\rangle$, under the constraint $|S_a-S_b| \leq n \leq S_a+S_b$.

Note that the preceding ground states are all unique, because in each case, not
only $S$ and $S_z$, but also $S_a$ and $S_b$ are specified.  It has
been known that for a single species of spin-$1$ atoms,
$|S_{\alpha},S_{\alpha}\rangle_{\alpha}$ is unique~\cite{hoyin}.
Hence any spin basis state
\begin{equation}
|S_{\alpha},S_{\alpha z}\rangle_{\alpha}=[r(S_{\alpha})\cdots r(S_{\alpha
z}+1)]^{-1} (S_{\alpha_-})^{S_{\alpha}-S_{\alpha
z}}|S_{\alpha},S_{\alpha}\rangle_{\alpha}, \label{low}
\end{equation}
is also unique, where
\begin{equation}
S_{\alpha_-} = \sqrt{2}(\alpha_1^{\dagger}\alpha_0 +
\alpha_0^{\dagger}\alpha_{-1})
\end{equation}
is the spin lowering operator,
\begin{equation}
r(m) \equiv \sqrt{(S+m)(S-m+1)}.
\end{equation}
Therefore any state $|S_a,S_b,S,S_z\rangle$  of a mixture of two species of spin-$1$ atoms, as can be obtained from the spin basis states of the two species, is unique.

\section{Ground states in terms of boson operators}

We now proceed to determine the expressions of these ground states
in terms of boson operators. First, it is straightforward to obtain
\begin{eqnarray}
|G\rangle_I & = & |N_a,N_b,N_a+N_b,N_a+N_b\rangle \\
& = & \frac{1}{\sqrt{N_a!N_b!}}(a_1^{\dagger})^{N_a}
(b_1^{\dagger})^{N_b}|0\rangle, \label{gi}
\end{eqnarray}
and
\begin{widetext}
\begin{eqnarray} |G\rangle_{II} &
= & |0,0,0,0\rangle \\
& = &
Z_{II}[2a_1^{\dagger}a_{-1}^{\dagger}-(a_0^{\dagger})^2]^{N_a/2}
[2b_1^{\dagger}b_{-1}^{\dagger}-(b_0^{\dagger})^2]^{N_b/2}|0\rangle,
\label{gii}
\end{eqnarray}
\end{widetext}
where
$Z_{II}=[(N_a/2)!(N_b/2)!2^{(N_a+N_b)/2}(N_a+1)!!(N_b+1)!!]^{1/2}$
is the normalization constant~\cite{ho2}. From this expression, it can be seen that $|G\rangle_I$ exists only if $N_a$ and $N_b$ are both even.

One can rewrite in terms of boson operators $|G\rangle_{I'}$, $|G\rangle_{III}$ and $|G\rangle_{IV}$, in which cases
$S_{\alpha}$ takes the largest possible value  $N_{\alpha}$, by
calculating the spin basis states of each species using (\ref{low}), with
\begin{equation}
|N_{\alpha},N_{\alpha}\rangle_{\alpha}
= \frac{1}{\sqrt{N_{\alpha}!}}{\alpha^{\dagger}}^{N_{\alpha}}|0\rangle,
\end{equation}
and then substituting them into (\ref{g12}), (\ref{g31}) and (\ref{g4}), respectively. For convenience in reading, we write them down explicitly in the following equations.
\begin{widetext}
\begin{eqnarray}
|G\rangle_{I'}
&=&|N_a,N_b,N_a+N_b,S_z\rangle\\
& =&
\sum_{S_{bz}=-N_b}^{N_b}
\frac{g_{I'}(S_{bz})2^{\frac{N_a+N_b-S_z}{2}}}{\sqrt{N_a!N_b!}
r(N_a)\cdots r(S_z-S_{bz}+1)r(N_b)\cdots r(S_{bz}+1) } \nonumber  \\
& & \times
(a_0^{\dagger}a_1+a_{-1}^{\dagger}a_0)^{N_a-S_z+S_{bz}}
{a_1^{\dagger}}^{N_a}
(b_0^{\dagger}b_1+b_{-1}^{\dagger}b_0)^{N_b-S_{bz}}
{b_1^{\dagger}}^{N_b} |0\rangle,
\end{eqnarray}
where
\begin{eqnarray}
g_{I'}(S_{bz})&=& \sqrt{ \frac{(2N_a)!(2N_b)!(N_a+N_b+S_z)!(N_a+N_b-S_z)! }
{(2N_a+2N_b)! (N_a-S_z+S_{bz})! (N_a+S_z-S_{bz})!
(N_b-S_{bz})!(N_b+S_{bz})!}}.
 \end{eqnarray}
\begin{eqnarray}
|G\rangle_{III} & = & |N,N,0,0\rangle   \\
\displaystyle  & =  & \frac{2^N}{(2N+1)N!}\sum_{m=-N}^N
\frac{(-1)^m}{
r(N)\cdots r(m+1)r(N)\cdots r(-m+1) } \nonumber  \\
& & \times
(a_0^{\dagger}a_1+a_{-1}^{\dagger}a_0)^{N-m}
{a_1^{\dagger}}^{N}
(b_0^{\dagger}b_1+b_{-1}^{\dagger}b_0)^{N+m} {b_1^{\dagger}}^{N} |0\rangle, \label{35}
\end{eqnarray}

\begin{eqnarray}
|G\rangle_{IV} & = & |N_a,N_b,n,n\rangle\\
& = & \sum_{S_{bz}=-N_b}^{N_b}
\frac{g_{IV}(S_{bz})2^{\frac{N_a+N_b-n}{2}}}{\sqrt{N_a!N_b!}
r(N_a)\cdots r(n-S_{bz}+1)r(N_b)\cdots r(S_{bz}+1) } \nonumber  \\
& & \times
(a_0^{\dagger}a_1+a_{-1}^{\dagger}a_0)^{N_a-n+S_{bz}}
{a_1^{\dagger}}^{N_a}
(b_0^{\dagger}b_1+b_{-1}^{\dagger}b_0)^{N_b-S_{bz}}
{b_1^{\dagger}}^{N_b} |0\rangle, \label{37}
\end{eqnarray}
where
\begin{eqnarray}
g_{IV}(S_{bz})&=& (-1)^{S_{bz}} \sqrt{ \frac{(N_a+N_b-n)!(2n+1)! (N_a+n-S_{bz})!(N_b+S_{bz})!}
{(N_a+N_b+n+1)! (n+N_a-N_b)!(n-N_a+N_b)! (N_a-n+S_{bz})!
(N_b-S_{bz})!}}. \label{38}
 \end{eqnarray}

\end{widetext}

$|S_b,S_b,0,0\rangle$, which is the ground state in some boundary regime of regime III, can be given by (\ref{35}), with $N$ replaced by $S_b$.  $|S_a,S_b,n,n\rangle$, which is the ground state in some boundary regime of regime IV, can be given by (\ref{37}) and (\ref{38}), with $N_a$ and $N_b$ replaced by $S_a$ and $S_b$, respectively.

A clearer picture of the composite structures of these ground states in terms of basic units is revealed by generalizing the method
of generating function~\cite{hoyin,koashi}, in the next section.

\section{ composite structures of the ground states}

\subsection{Generating function method for
$|S_a,S_b,S,S\rangle$}

Now we consider the construction of $|S_a,S_b,S, S\rangle$ of a mixture of two
species of spin-$f$ atoms, with total spin and its $z$-component both being $S$.
With $S_z$ being maximal, we may consider a configuration of the state $|S_a,S_b,S, S\rangle$, in which there are
$Q_{m_j,n_j,l_j}$ copies of unit $j$, which is made up of $m_j$ $a$-atoms and $n_j$
$b$-atoms and carrying spin $l_j$. Denoting  the creation operators for unit $j$ as
$\Theta_{m_j,n_j,l_j}^{\dagger}$, we have
\begin{widetext}
\begin{equation}
|S_a,S_b, S,S\rangle = \displaystyle \sum
A(\{Q_{m_j,n_j,l_j}\}) \prod (\Theta_{m_j,n_j,l_j}^{\dagger})^{
Q_{m_j,n_j,l_j}}|0\rangle, \end{equation}
\end{widetext}
where $A$ is a coefficient and the summation is over all possible values of
$\{Q_{m_j,n_j,l_j}\}$.

The possible values of $m_j$, $n_j$, $l_j$ and $Q_{m_j,n_j,l_j}$ are subject to the  constraints from $N_a$, $N_b$ and $S$,
\begin{equation}
\begin{array}{rcl}
\displaystyle \sum_{j} n_j Q_{m_j,n_j,l_j} & = & N_a,\\
\displaystyle \sum_{j} m_j Q_{m_j,n_j,l_j} & = & N_b, \\
\displaystyle \sum_{j} l_j Q_{m_j,n_j,l_j} & = & S,
\end{array}
\end{equation}
as well as the constraints from  $S_a$ and $S_b$,
\begin{equation}
\begin{array}{rcl}
S_a^2  |S_a,S_b, S,S\rangle & = & S_a(S_a+1)  |S_a,S_b, S,S\rangle, \\
S_b^2  |S_a,S_b, S,S\rangle & = & S_b(S_b+1)  |S_a,S_b, S,S\rangle.
\end{array} \label{sab}
\end{equation}
where $S_{\alpha}^2 = (\alpha_1^{\dagger}\alpha_1- \alpha_{-1}^{\dagger}\alpha_{-1})^2
+ 2(\alpha_0^{\dagger})^2\alpha_1\alpha_{-1} + 2 \alpha_1^{\dagger}\alpha_{-1}^{\dagger}\alpha_0^2 + 2\alpha_0^{\dagger} \alpha_0\alpha_1\alpha_1^{\dagger} + 2 \alpha_0^{\dagger} \alpha_0\alpha_{-1}\alpha_{-1}^{\dagger} $, as one can easily find.

For an integer $f$, we define the generating function as
\begin{equation}
G(x_a,x_b,y) \equiv
\sum_{N_a,N_b,S}M(N_a,N_b,S)x_a^{N_a}x_b^{N_b}y^S, \label{gen}
\end{equation}
where $x_a$, $x_b$ and $y$ are complex numbers inside the unit
circle and $M(N_a,N_b,S)$ is the number of solutions of the sets of the
nonnegative integers $\{Q_{m_j,n_j,l_j}\}$.
Following the method of \cite{hoyin}, we obtain that
\begin{equation}
\begin{array}{l}
G(x_a,x_b,y) = \\
\displaystyle \int_{\cal C} \frac{dz}{2\pi
i} \frac{1-z^{-1}}{z-y} \prod_{j_a=-f}^{f}\prod_{j_b=-f}^{f} \frac{1}{(1-x_a
z^{j_a})(1-x_bz^{j_b})},
\end{array}
\end{equation}
where the contour integral is along the unit circle ${\cal C}$.

For $f=1$, we obtain that
\begin{widetext}
\begin{equation}
\begin{array}{rcl}
G(x_a,x_b,y) & = &
\sum
(x_a^{Q_{1,1,0}+2Q_{2,0,0}+Q_{1,0,1}}
x_b^{Q_{1,1,0}+2Q_{0,2,0}+Q_{0,1,1}} y^{Q_{1,0,1}+Q_{0,1,1}} \\
&& + \sum x_a^{Q_{1,1,0}+2Q_{2,0,0}+Q_{1,0,1}+1}
x_b^{Q_{1,1,0}+2Q_{0,2,0}+Q_{0,1,1}+1} y^{Q_{1,0,1}+Q_{0,1,1}+1}).
\end{array}
\label{gen2}
\end{equation}
\end{widetext}
where the summations are over all possible values of $Q_{1,1,0}$, $Q_{2,0,0}$, $Q_{0,2,0}$, $Q_{1,0,1}$ and $Q_{0,1,1}$, all of which are nonnegative.
Comparing (\ref{gen}) and (\ref{gen2}), we have
\begin{equation}
M(N_a,N_b,S)=M_1(N_a,N_b,S)+M_2(N_a,N_b,S),
\end{equation}
where
$M_1(N_a,N_b,S)$ is the number of solutions to the set of equations
\begin{equation}
\begin{array}{rcl}
Q_{1,1,0}+2Q_{2,0,0}+Q_{1,0,1}&=& N_a, \\
Q_{1,1,0}+2Q_{0,2,0}+Q_{0,1,1}&=& N_b, \\
Q_{1,0,1}+Q_{0,1,1}&=&S,
\end{array} \label{n1}
\end{equation}
while $M_2(N_a,N_b,S)$ is the number of solutions to the set of
equations
\begin{equation}
\begin{array}{rcl}
Q_{1,1,0}+2Q_{2,0,0}+Q_{1,0,1}+1&=& N_a, \\
Q_{1,1,0}+2Q_{0,2,0}+Q_{0,1,1}+1&=& N_b, \\
Q_{1,0,1}+Q_{0,1,1}+1&=&S.
\end{array} \label{n2}
\end{equation}

In general,  there may be multiple solutions to (\ref{n1}). In each
solution, there are $Q_{1,1,0}$ interspecies singlets consisting
of one $a$-atom and one $b$-atom, $Q_{2,0,0}$ singlets consisting
of two $a$-atoms and $Q_{0,2,0}$ singlets consisting of two
$b$-atoms. In addition, there are $Q_{1,0,1}$ $a$-atoms with
$z$-component spin $1$, as well as $Q_{0,1,1}$ $b$-atoms
with $z$-component spin $1$.

There may also be multiple solutions to (\ref{n2}). In each
solution, there are $Q_{1,1,0}$ interspecies singlets consisting
of one $a$-atom and one $b$-atom, $Q_{2,0,0}$ singlets consisting
of two $a$-atoms and $Q_{0,2,0}$ singlets consisting of two
$b$-atoms. Also, either there are $Q_{1,0,1}$ $a$-atoms
with $z$-component spin $1$ together with one $a$-atom with
z-component spin $0$, as well as $Q_{0,1,1}+1$ $b$-atoms
with $z$-component spin $1$; or there are $Q_{0,1,1}$
$b$-atoms with $z$-component spin $1$ together with  one $b$-atom
with z-component spin $0$, as well as $Q_{1,0,1}+1$
$a$-atoms with $z$-component spin $1$.

\subsection{$|G\rangle_{I}$ and $|G\rangle_{II}$}

As a simple example, we can verify that $|G\rangle_I=|N_a,N_b,N_a+N_b\rangle = |N_{a}, N_a\rangle_a\otimes |N_b, N_b\rangle_b$ is indeed as given in (\ref{gi}). For $S=N_a+N_b$, the only solution to (\ref{n1}) is $Q_{1,0,1}=N_a$, $Q_{0,1,1}=N_b$, and $Q_{1,1,0}=Q_{2,0,0}=Q_{0,2,0}=0$, while there is no solution to (\ref{n2}). Hence in this case, $|G\rangle_I$ must be given by (\ref{gi}), which is obviously an eigenstate of $S_a^2$ and $S_b^2$, as it should be.

For $S=0$, there is no solution to (\ref{n2}), while there are solutions to (\ref{n1}) with   $Q_{1,0,1}=Q_{0,1,1}=0$,
$Q_{2,0,0}=(N_a-Q_{1,1,0})/2$, and
$Q_{0,2,0}=(N_b-Q_{1,1,0})/2$.
Hence a state $|S_a,S_b,0,0\rangle$ can be expressed as
\begin{widetext}
 \begin{equation}
 |S_a,S_b,0,0\rangle = \sum_{Q_{1,1,0}} A(Q_{1,1,0})  (\Theta_{1,1,0}^{\dagger})^{Q_{1,1,0}} (\Theta_{2,0,0}^{\dagger})^{(N_a-Q_{1,1,0})/2}
(\Theta_{0,2,0}^{\dagger})^{(N_b-Q_{1,1,0})/2}|0\rangle, \label{exp}
\end{equation}
\end{widetext}
where
\begin{equation}
{\Theta_{1,1,0}}^{\dagger}=a_1^{\dagger}b_{-1}^{\dagger}
-a_0^{\dagger}b_{0}^{\dagger}+a_{-1}^{\dagger}b_{1}^{\dagger}
\end{equation}
is the creation operator for an interspecies singlet pair, while $\Theta_{2,0,0}^{\dagger} = 2a_1^{\dagger}a_{-1}^{\dagger}
-{a_0^{\dagger}}^2$ and $\Theta_{0,2,0}^{\dagger} = 2b_1^{\dagger}b_{-1}^{\dagger}
-{b_0^{\dagger}}^2$ are creation operators of intraspecies singlet pairs of $a$-atoms and $b$-atoms, respectively.
$ |S_a,S_b,0,0\rangle$ is a superposition of configurations with all possible values of $Q_{1,1,0}$, in each of which there are $Q_{1,1,0}$ interspecies singlet pairs, $(N_a-Q_{1,1,0})/2$ singlet pairs of two $a$-atoms and  $(N_b-Q_{1,1,0})/2$ singlet pairs of two $b$-atoms. In addition, (\ref{sab}) with $S_a=S_b=N$ imposes constraints determining $C(Q_{1,1,0})$.

Equation (\ref{gii2}) indicates that in  $|G\rangle_{II} =|0,0\rangle_a\otimes |0,0\rangle_b$, the two species are disentangled. Therefore $Q_{1,1,0}=0$, and thus
$Q_{2,0,0}=N_a/2$, and $Q_{0,2,0} = N_b/2$; that is, there are only
intraspecies singlet pairs. Hence (\ref{gii}) is confirmed. As mentioned previously, this is subject to the condition that $N_a$ and $N_b$ are both even.

Similarly, we would like to note that for a gas of a single species $\alpha$ of $N$ spin-$1$ atoms with $c^{\alpha} > 2\gamma B$, if $N$
is odd, the (non-normalized) ground state should be $|1,1\rangle =
\alpha_1^{\dagger}(2\alpha_1^{\dagger}\alpha_{-1}^{\dagger}
-{\alpha_0^{\dagger}}^2)^{(N-1)/2}|0\rangle$, rather than the singlet $|0,0\rangle$, which exists only for even $N$.

\subsection{$|G\rangle_{III}$ and $|G\rangle_{IV}$}

Now we look at $|G\rangle_{III} =  |N,N,0,0\rangle$. Equation (\ref{g31}) already indicates that the two species are strongly entangled, therefore $|G\rangle_{III}$ is in the form of (\ref{exp}).

For $|G\rangle_{IV}$, one considers (\ref{n1}) and (\ref{n2}) with $S=S_z=n$.  It can be found that the solution to (\ref{n1}) is
\begin{equation}
\displaystyle
\left\{ \begin{array}{lcl}
Q_{1,0,1} & = & \frac{N_a-N_b+n}{2} + Q_{2,0,0} - Q_{0,2,0}, \\
Q_{0,1,1} & = & \frac{N_b-N_a+n}{2} + Q_{0,2,0} - Q_{2,0,0},
\end{array} \right. \label{q1}
\end{equation}
which is valid if $N_a+N_b-n$ is even,
while the solution to (\ref{n2}) is
\begin{equation}
\displaystyle
\left\{ \begin{array}{lcl}
Q_{1,0,1} & = & \frac{N_a-N_b+n-1}{2} + Q_{2,0,0} - Q_{0,2,0}, \\
Q_{0,1,1} & = & \frac{N_b-N_a+n-1}{2} + Q_{0,2,0} - Q_{2,0,0}. \end{array} \right. \label{q2}
\end{equation}
which is valid if $N_a+N_b-n$ is odd.

Hence, if $N_a+N_b-n$ is even,
\begin{widetext}
\begin{equation}
|G\rangle_{IV}=|N_a,N_b,n,n\rangle= \sum
A(Q_{1,1,0},Q_{2,0,0},Q_{0,2,0})
 {a_1^{\dagger}}^{Q_{1,0,1}}
{b_1^{\dagger}}^{Q_{0,1,1}}
{\Theta_{1,1,0}^{\dagger}}^{Q_{1,1,0}}
{\Theta_{2,0,0}^{\dagger}}^{Q_{2,0,0}} {\Theta_{0,2,0}^{\dagger}}^{Q_{0,2,0}}|0\rangle,
\label{g42}
\end{equation}
where $Q_{1,0,1}$ and $Q_{0,1,1}$ are given by (\ref{q1}) while
if $N_a+N_b-n$ is odd,
\begin{equation}
\begin{array}{cl}
\displaystyle
|G\rangle_{IV}=|N_a,N_b,n,n\rangle = &
\sum
A(Q_{1,1,0},Q_{2,0,0},Q_{0,2,0}) a_0^{\dagger}
 {a_1^{\dagger}}^{Q_{1,0,1}}
{b_1^{\dagger}}^{Q_{0,1,1}+1}
{\Theta_{1,1,0}^{\dagger}}^{Q_{1,1,0}}
{\Theta_{2,0,0}^{\dagger}}^{Q_{2,0,0}} {\Theta_{0,2,0}^{\dagger}}^{Q_{0,2,0}}|0\rangle \\
\displaystyle
&
+\sum A'(Q_{1,1,0},Q_{2,0,0},Q_{0,2,0}) {a_1^{\dagger}}^{Q_{1,0,1}+1}b_0^{\dagger}
{b_1^{\dagger}}^{Q_{0,1,1}}
{\Theta_{1,1,0}^{\dagger}}^{Q_{1,1,0}}
{\Theta_{2,0,0}^{\dagger}}^{Q_{2,0,0}} {\Theta_{0,2,0}^{\dagger}}^{Q_{0,2,0}}|0\rangle,
\end{array}
\label{g422}
\end{equation}
\end{widetext}
  where $Q_{1,0,1}$ and $Q_{0,1,1}$ are given by (\ref{q2}). In both (\ref{g42}) and (\ref{g422}),
 the summations  are  over $Q_{1,1,0},$ $Q_{2,0,0}$ and $Q_{0,2,0}$.
 The coefficients $A$ and $A'$ are determined by constraints (\ref{sab}) with $S_a=N_a$ and $S_b=N_b$.

\section{ composite structures of $|S,S\rangle$ of a pseudospin-$1/2$ mixture}

As a comparison, we now apply the generating function method to determine $|S,S\rangle$ of a mixture of two species of pseudospin-$1/2$ atoms. Unlike the case of spin-1, for
each species $\alpha$ of pseudospin-$\frac{1}{2}$, the total spin is
always fixed to be $S_{\alpha}=N_{\alpha}/2$. With $N_a$ and $N_b$ fixed, the spin state of the pseudospin-$\frac{1}{2}$ mixture is only determined by $S$ and $S_z$ of the total spin; that is,
\begin{equation}
|S,S_z\rangle \equiv |\frac{1}{2}N_a, \frac{1}{2}N_b, S,S_z\rangle.
\end{equation}

The basic method for constructing the maximally polarized $|S,S\rangle$ remains the same as that for a spin-1 mixture, as described in Sec.~V.A. In a configuration of the state $|S, S\rangle$, there are
$Q_{m_j,n_j,l_j}$ copies of unit j, which is made up of $m_j$ $a$-atoms and $n_j$
$b$-atoms and carries spin $l_j/2$.  The generating function is now defined as
\begin{equation}
G(x_a,x_b,y) \equiv
\sum_{N_a,N_b,S}M(N_a,N_b,S)x_a^{N_a}x_b^{N_b}y^{2S}.
\end{equation}
It can be obtained that
\begin{equation}
\begin{array}{l}
G(x_a,x_b,y) = \\
\displaystyle \int_{\cal C} \frac{dz}{2\pi
i} \frac{1-z^{-2}}{z-y}  \prod_{j_a=-f}^{f}\prod_{j_b=-f}^{f}\frac{1}{(1-x_a
z^{2j_a})(1-x_bz^{2j_b})},
\end{array}
\end{equation}
where the contour integral is along the unit circle ${\cal C}$. Then one obtains
\begin{equation}
G(x_a,x_b,y) = \sum
x_a^{Q_{1,1,0}+Q_{1,0,1}} x_b^{Q_{1,1,0}+ Q_{0,1,1} }y^{Q_{1,0,1}+Q_{0,1,1}}.
\label{gen13}
\end{equation}
Therefore,
\begin{equation}
\begin{array}{rcl}
Q_{1,1,0}+ Q_{1,0,1}&=& N_a, \\
Q_{1,1,0}+ Q_{0,1,1} &=& N_b, \\
Q_{1,0,1}+Q_{0,1,1} &=&2S,
\end{array} \label{n12}
\end{equation}
which has a unique solution,
\begin{equation}
\begin{array}{rcl}
Q_{1,0,1}&=&S+\frac{1}{2}(N_a-N_b),  \\
Q_{0,1,1}&=&S+\frac{1}{2}(N_b-N_a), \\
Q_{1,1,0}&=&S+\frac{1}{2}(N_a+N_b).
\end{array} \label{n13}
\end{equation}
which is subject to the condition that $Q_{1,1,0}$,
$Q_{1,0,1}$ and $Q_{0,1,1}$ should all be nonnegative integers. Under
this condition,
\begin{widetext}
\begin{equation}
|S,S\rangle =
{a_{\uparrow}^{\dagger}}^{S+(N_a-N_b)/2}{b_{\uparrow}^{\dagger}}^{S+(N_b-N_a)/2}
{{\Phi_{1,1,0}}^{\dagger}}^{S+(N_a+N_b)/2}|0\rangle,
\end{equation}
\end{widetext}
where the normalization constant is neglected, and
\begin{equation}
{\Phi_{1,1,0}}^{\dagger}=a_{\uparrow}^{\dagger}b_{\downarrow}^{\dagger} -a_{\downarrow}^{\dagger}b_{\uparrow}^{\dagger}
\end{equation}
is the interspecies singlet creation operator for a pseudospin-$\frac{1}{2}$ mixture. A sufficient and necessary condition for $Q_{1,1,0}$, $Q_{1,0,1}$ and $Q_{0,1,1}$ all to be integers is that $2S+N_a+N_b$ is an even integer.

When $S=0$, the only consistent solution of (\ref{n13}) is
$Q_{1,0,1}=Q_{0,1,1}=0$ while $Q_{1,1,0}=N$ if and only if
$N_a=N_b=N$. The state is then the global singlet
\begin{equation}
|0,0\rangle =
{{\Phi_{1,1,0}}^{\dagger}}^N|0\rangle.  \label{single}
\end{equation}
When $N_a \geq N_b$, the smallest value of $S$ is $(N_a-N_b)/2$, for which $Q_{1,1,0}=N_b$,
$Q_{1,0,1}=N_a-N_b$, and $Q_{0,1,1}=0$,  hence
\begin{widetext}
\begin{equation}
|\frac{1}{2}(N_a-N_b),\frac{1}{2}(N_a-N_b)\rangle =
{a_{\uparrow}^{\dagger}}^{N_a-N_b}
{{\Phi_{1,1,0}}^{\dagger}}^{N_b}|0\rangle,  \label{ab}
\end{equation}
\end{widetext}
of which the singlet state (\ref{single}) is a special case.

\section{Summary}

\begin{table}
\begin{tabular}{|l|l|l|}
\hline \multirow{2}{*}{No.} &  Parameter  & \multirow{2}{*}{Ground states} \\
 & regimes & \\
 \hline
 \multirow{3}{*}{I} &  $c^{ab} \leq 0$, &  $|G\rangle_I=|N_a,N_b,N_a+N_b,N_a+N_b\rangle$  \\
 & $c^{a} < c^{ab}$,  & $= |N_a,N_a\rangle_a \otimes |N_b,N_b\rangle_b$  \\ & $c^{b} < c^{ab}$. & (disentangled)\\
 \hline
 \multirow{3}{*}{II} &  $c^{ab} \geq 2\gamma B $, &  $|G\rangle_{II}=|0,0,0,0\rangle$  \\
 & $c^{a} > c^{ab}$,  & $= |0,0\rangle_a \otimes |0,0\rangle_b$   \\ & $c^{b} > c^{ab}$. & (disentangled)\\
 \hline
 \multirow{4}{*}{III} &  $c^{ab} \geq  2\gamma B >0 $ &  $|G\rangle_{III} =|N,N,0,0\rangle$  \\
 & or  $c^{ab} > 2\gamma B = 0 $,   &  (entangled) \\ &
 $c^{a} < c^{ab}$, &  \\ &
  $c^{b} < c^{ab}$  &  \\&
 $N_a=N_b=N$. &  \\
 \hline
 \multirow{3}{*}{IV} &  $ 0 < c^{ab} \leq   2\gamma B $, &  $|G\rangle_{IV}= |N_a,N_b,n,n\rangle$  \\
 & $c^{a} < c^{ab}$,  &  $n \equiv Int(\frac{\gamma B
}{c^{ab} }- \frac{1}{2})$    \\ & $c^{b} < c^{ab}$. & (entangled)   \\
 \hline
 \end{tabular}
\caption{Ground stats of a mixture of  two spin-1 atomic gases in four parameter regimes. The first (I) is a direct product of ferromagnetic states of independent BEC of the two species. The second (II) is a direct product of the two independent singlet states of the two species. The third (III) is an interspecies singlet state, which we call EBEC. The fourth (IV) is also EBEC if $n<N_a+N_b$.Exactly speaking, $|G\rangle_{II}$ is subject to the condition that $N_a$ and $N_b$ are both even.  Situations on the boundaries of these parameter regimes  are discussed in the text. }
\end{table}

To summarize, we have considered a mixture of two different species of spin-1 gases with interspecies spin-exchange scattering, as an extension of our previous work on a mixture of two different species of pseudospin-$\frac{1}{2}$ gases, going beyond the usual spinor  bose gases and BEC mixtures without interspecies entanglement. Interspecies spin exchange favors spin ordering between different species, while intraspecies spin exchange favors spin ordering within each species. The ground state of such a  mixture thus depends on the parameters in the many-body Hamiltonian, which we have shown to be reduced to a Hamiltonian of two giant spins.

We have worked out the ground states in four typical parameter regimes, which are now reported in Table~I. It is straightforward to verify that they are all fragmented BEC, by calculating one-particle reduced density matrices. When $c^{a}$ and
$c^{b}$ are both less than $c^{ab}$, which is negative or $0$, all atoms of each species form a ferromagnetic state with the spin of each atom being $\mu=1$; that is,  the ground state of the mixture is the direct product of two independent ferromagnetic states.  When $c^{a}$ and
$c^{b}$ are both larger than $c^{ab}$, which is larger than or equal to $2\gamma B$, the atoms of each species form a singlet state, with the total spin of each species being $0$; that is,  the ground state of the mixture is the direct product of  two independent singlet states, subject to the condition that $N_a$ and $N_b$ are even (otherwise there is minute deviation).  These two ground states are disentangled between the two species.   For $N_a=N_b=N$, when $c^{ab} \geq
 2 \gamma B > 0$ or $c^{ab} >
 2 \gamma B =0$, while $c^{a}$ and $c^{b}$ are both less than $c^{ab}$, the ground state is a global singlet state $|G\rangle_{III} =|N,N,0,0\rangle$, with total spin zero. When $0< c^{ab} \leq 2\gamma B  $,and  $c^{a}$ and $c^{b}$ are both less than $c^{ab}<0$, the ground state is $|G\rangle_{IV}= |N_a,N_b,n,n\rangle$, where $n \equiv Int(\frac{\gamma B
}{c^{ab} }- \frac{1}{2})$.

The latter two ground states exhibit  EBEC,  displaying strong interspecies entanglement. A consequence of this entanglement is that the particle number in each spin state of each species is subject to strong quantum fluctuation.  There are rich composite structures due to interspecies entanglement. By using the generating function method, it has been revealed that $|G\rangle_{III}$ and $|G\rangle_{IV}$ are each superpositions of configurations with both intraspecies and interspecies singlet pairs.
It is interesting to note that $|G\rangle_{III}=|G\rangle_{IV}$ when $c^{ab}=2\gamma B$, $N_a=N_b$, implying that $|G\rangle_{III}$  and $|G\rangle_{IV}$ belong to the same quantum phase.

We have also used the generating function method to find the spin state $|S,S\rangle$ of a mixture of two species of pseudospin-$\frac{1}{2}$ atoms with interspecies spin exchange. As the total spin of each species of pseudospin-$\frac{1}{2}$ atoms is always half of the atom number, the  composite structure of a  pseudospin-$\frac{1}{2}$ atoms is simpler, with only one configuration. Consequently, there are only interspecies singlet pairs when the total spin of the mixture is zero, hence EBEC in such a case is simply BEC occurring in an interspecies singlet state. Such a simplicity is lost in a spin-$1$ mixture, in which intraspecies singlet pairs coexist with interspecies singlet pairs, and  EBEC is generally defined as BEC with interspecies entanglement. Previous studies on pseudospin-$\frac{1}{2}$ mixture demonstrated that  EBEC leads to various physical properties different from those of usual BEC,  to be similarly studied in spin-$1$ mixtures.

\acknowledgments

I am grateful to my student Li Ge for noting that ${\Theta_{1,1,0}^{\dagger}}^N|0\rangle$,  written as ${{\Theta^{(ab)}_{1,1}}^{\dagger}}^N|0\rangle$ in an earlier version of this paper~\cite{present}, is not an eigenstate of $S_a^2$ or $S_b^2$. I also thank Jason Ho and Lan Yin for brief discussions.  This work was supported by the National Science Foundation of China (Grant No. 10674030), the Shuguang Project (Grant No. 07S402) and the Ministry of Science and Technology of China (Grant No. 2009CB929204).

{\em Note added:}  Recently we became aware of a paper \cite{xu2} that simply gives $|N_a,N_b,N_a-N_b,S_z\rangle$ as the ground state ``for large antiferromagnetic spin-exchange interaction between the two species'', without actually showing it gives the lowest energy. We had determined that the ground state varies with the parameters, and is not always this state even for large antiferromagnetic interspecies spin-exchange interaction. They also applied the same generating function method  as ours to the Hamiltonian with an additional $P_0$ term, which, to our understanding, had been shown to vanish by Luo {\it et al.}~\cite{ma}. We also disagree with their statement that there is no interspecies singlet pairing when this additional $P_0$ term vanishes. We have worked out all possible ground states, which will be discussed elsewhere

\end{document}